\documentclass[aip,preprint]{revtex4-1}
\usepackage{graphicx}

\usepackage{appendix}
\usepackage{amsmath}
\usepackage{comment}
\usepackage{color}

\begin{document}

\title{Path Integral Monte Carlo for Fictitious Identical Particles with $\xi$-Ensemble}

\author{Yunuo Xiong}
\email{xiongyunuo@hbpu.edu.cn}
\affiliation{Center for Fundamental Physics, Hubei Polytechnic University, Huangshi 435003, China}
\date{\today}

\begin{abstract}
In this work, a path integral Monte Carlo (PIMC) algorithm for fictitious identical particles (FIP) is proposed by introducing a $\xi$-ensemble and performing PIMC simulations on the resulting $\xi$-ensemble partition function. The PIMC algorithm with the $\xi$-ensemble allows us to obtain the thermodynamic properties of FIP for different $\xi$ values in a single simulation. Moreover, it also accelerates the simulation by improving the sampling efficiency, compared to the usual case where independent simulations are performed for each $\xi$ value, in the sense that the autocorrelation time of samples belonging to the same $\xi$ sector is decreased. Simulations of the uniform electron gas and uniform warm dense beryllium are performed to validate the improved algorithm and study the improvement in its sampling efficiency.
\end{abstract}

\maketitle

\section{Introduction}
Path integral Monte Carlo (PIMC) \cite{herman1982on,minoru1984monte,ceperley1995path} is a powerful method for studying the thermodynamic properties of quantum many-body systems. However, when applied to fermionic systems, it suffers from the well-known fermion sign problem \cite{Ceperley1980,alexandru2022complex,Zi-Xiang,Shiwei,Ali,Boris,hirshberg2020path,dornheim2020attenuating,dornheim2019fermion,dornheim2021fermion,xiong2025pseudo,YunuoUEG,yu2026preempting,LiDeep,qiao2026neural,Cataldo}. On the other hand, the fictitious identical particles (FIP) method \cite{xiong2022thermodynamic,xiong2023thermodynamics,dornheim2023fermionic} has emerged as a promising approach for circumventing the fermion sign problem by introducing a parameter, $\xi$, that characterizes fictitious quantum statistics and continuously interpolates between Bose and Fermi statistics. Thermodynamic properties of fermions are then obtained via $\xi$-extrapolation from the bosonic sector, where the sign problem is absent. The FIP method has found successful applications in a number of large-scale fermionic systems, including warm dense matter \cite{dornheim2023fermionic,dornheim2024ab,dornheim2025unraveling,dornheim2026taylor,dornheim2025fermionic,vorberger2025roadmap,bonitz2024toward,dornheim2026overview}, helium-3 \cite{morresi2025normal}, the strongly degenerate uniform electron gas \cite{morresi2025study}, and the Fermi-Hubbard model \cite{fan2025quantum}. There has also been an intriguing theoretical investigation by Li’s group into the properties of the FIP partition function from the perspective of Lee–Yang $\xi$-zeros \cite{he2025revisiting,he2026fermion}.
\par
Generally speaking, in order to apply FIP method in practice, we need a decent number of simulation data in the bosonic sector. The worm algorithm \cite{prokof2006worm} is an efficient implementation of PIMC for sampling bosonic distributions, and has been extended by Dornheim et al. \cite{dornheim2023fermionic} to incorporate permutation statistics of FIP. However, in the traditional worm algorithm implementation, in order to calculate thermodynamic properties for different $\xi$ values, independent PIMC simulations must be carried out, one for each $\xi$ value. This scheme may not be efficient enough when simulation data for a large number of $\xi$ values are needed, or when the system size is large. So in this work, an improved version of worm algorithm for FIP is proposed by treating the sampling distribution as an ensemble of different $\xi$, so that in one single PIMC simulation with $\xi$-ensemble we can obtain properties for many different $\xi$ values in the bosonic sector, and the autocorrelation time for samples corresponding to the same $\xi$ sector is reduced allowing for greater sampling efficiency as well. We believe the PIMC algorithm in this work can be employed to accelerate FIP simulation for existing and new applications, some results for the simulation of uniform electron gas and uniform beryllium gas based on FIP are shown to support the claim about efficiency in this work.
\par
This paper is organized as follows. In Sec. II we give a brief review on FIP and worm algorithm. In Sec. III the methodology for PIMC based on $\xi$-ensemble will be introduced in details, and in Sec. IV simulation results for uniform electron gas and uniform beryllium gas are presented to validate the method and analyze its sampling efficiency by calculating autocorrelation time. Finally a summary is given in Sec. V.

\section{Background}
In the FIP framework\cite{xiong2022thermodynamic,xiong2023thermodynamics,dornheim2023fermionic}, the partition function describing its thermodynamics can be written using continuous path integral formalism as follows (for the fully polarized system)
\begin{equation}
Z(\beta,\xi) =  \frac{1}{N!}\sum_{p \in S_N} \xi^{N_p} \int_{\mathbf{x}_i(\hbar\beta)=\mathbf{x}_{p(i)}(0)} \mathcal{D}[\mathbf{x}(\tau)] \exp \left( -\frac{1}{\hbar} \int_0^{\hbar\beta} L_E d\tau \right),
\end{equation}
where $p$ is the permutation operator for $N$ particles, and $p(i)$ is the permutated particle index, $N_p$ is the number of exchanges in the permutation. $\mathbf{x}(\tau)$ represents the particles' continuous path in imaginary time and $L_E$ is the Euclidean Lagrangian given by
\begin{equation}
L_E(\mathbf{x}(\tau),\dot{\mathbf{x}}(\tau))=\frac{m}{2}\dot{\mathbf{x}}(\tau)^2+V(\mathbf{x}(\tau)),
\end{equation}
where $V$ is the potential function. The parameter $\xi$ characterizes FIP: $\xi=1$ corresponds to bosons, $\xi=-1$ corresponds to fermions, while for other real $\xi$ values inside $[-1,1]$ it describes fictitious quantum statistics between bosons and fermions. In the region where $\xi\geq 0$ (called bosonic sector), the weight factor as defined by $Z(\beta,\xi)$ is positive definite and can be interpreted as a probability distribution, and Monte Carlo sampling can be directly applied free from fermion sign problem.
\par

$\xi$-extrapolation\cite{xiong2022thermodynamic,xiong2023thermodynamics,dornheim2023fermionic} works by performing simulations in the bosonic sector and extrapolate to the fermionic sector. For example, we may treat the expectation value of an observable $\langle O\rangle$ as a function of $\xi$, $\langle O(\xi)\rangle$, at fixed temperature $T$. Assuming that $\langle O(\xi)\rangle$ is an analytical function of $\xi$, we can use a polynomial function to approximate it and extrapolate to the fermionic sector in order to obtain properties for fermions. This so called isothermal extrapolation\cite{xiong2022thermodynamic,dornheim2023fermionic} has been successfully applied to weakly degenerate quantum systems\cite{dornheim2023fermionic,dornheim2024ab,dornheim2025unraveling,dornheim2026taylor,dornheim2025fermionic} (e.g. the regime where the temperature is high compared to Fermi energy).
\par
To perform PIMC simulations in the bosonic sector, we use a variant of the worm algorithm to include the exchange statistics of FIP. First, we have to discretize the continuous path integral expression. Using a 2nd order expansion for the propagator, it can be shown that\cite{spada2022path} the partition function for a system with periodic boundary condition is now given by
\begin{equation}
\begin{split}
Z(\beta,\xi)=\lim_{P\rightarrow\infty}(\frac{mP}{2\pi\beta\hbar^2})^{PdN/2}\frac{1}{N!}\sum_{p,\mathbf{W}}\xi^{N_p}\int d\mathbf{R}_1...d\mathbf{R}_P\\
e^{-\Delta\beta\sum_{k=1}^PU(\mathbf{R}_k)-\frac{mP}{2\beta\hbar^2}[(\mathbf{R}_P-p\{\mathbf{R}_{1}\}+\mathbf{W}L)^2+\sum_{k=1}^{P-1}(\mathbf{R}_k-\mathbf{R}_{k+1})^2]},
\end{split}
\end{equation}
where $d$ is the dimensionality, $\mathbf{R}_1$ to $\mathbf{R}_P$ are the particle coordinates at each imaginary time slice, and $p\{\mathbf{R}_{1}\}$ is the permutated particle coordinates. $L$ is the box length and $\mathbf{W}$ is the integer vector of winding numbers. $\Delta\beta=\beta/P$ and $U$ is the potential function. In practice, we choose a finite number of imaginary time slices $P$ to ensure convergence.
\par
The joint probability distribution that we want to sample is given by
\begin{equation}
\text{Pr}(\mathbf{R}_1,...,\mathbf{R}_P,p,\mathbf{W})=\xi^{N_p}e^{-\Delta\beta\sum_{k=1}^PU(\mathbf{R}_k)-\frac{mP}{2\beta\hbar^2}[(\mathbf{R}_P-p\{\mathbf{R}_{1}\}+\mathbf{W}L)^2+\sum_{k=1}^{P-1}(\mathbf{R}_k-\mathbf{R}_{k+1})^2]}.
\end{equation}
The worm algorithm uses Markov Chain Monte Carlo to sample the above distribution. In particular, the worm algorithm in Ref.\cite{spada2022path} specifies how to perform the simulation for periodic box, and it is also the implementation used in this work. Since $\xi$ enters only through the prefactor $\xi^{N_p}$ that is related to the current permutation $p$ only, any Monte Carlo moves that do not change the permutation are not affected by the inclusion of FIP. In standard worm algorithm implementation, the only move that changes permutation is the Swap move, so the acceptance probability for that move needs to be modified as
\begin{equation}
A_{swap,\xi}=\min\{1,\xi^{N_{p'}-N_p}\times A_{swap}\},
\end{equation}
where $p'$ is the new permutation and $p$ is the old permutation. Explicit formula of the original $A_{swap}$ and acceptance probabilities for other Monte Carlo moves are given in Ref.\cite{spada2022path}.

\section{PIMC with $\xi$-ensemble}
In the standard PIMC algorithm presented in the last section, $\xi$ is a fixed system parameter. So to obtain thermodynamic properties for different $\xi$, independent PIMC simulations must be carried out. A more efficient approach to the simulation of FIP can be considered by selecting a set of discrete $\xi$ values, $\{\xi_i\}$, in the bosonic sector (i.e. inside the region $[0,1]$), and then defining the $\xi$-ensemble partition function as
\begin{equation}
Z_{ensemble}(\beta)=\sum_ia_iZ(\beta,\xi_i),
\label{Zen}
\end{equation}
where $a_i$ are constants controlling the ration of the partition functions in each $\xi_i$ sector, and they are yet to be determined from actual PIMC implementation. Now, in order to perform PIMC simulation for the $\xi$-ensemble partition function, we need to specify which $\xi$-sector the current simulation is in, which is done by including $\xi_i$ as part of the sample. While all the Monte Carlo moves for FIP are not changed when considering $\xi$-ensemble, we just have to add a new type of Monte Carlo move to transition from one $\xi$-sector to another. The $\xi$-sector changing move is defined as follows:
\par
\textit{$\xi$-Translate}. If the current state is in the $\xi_i$-sector, a Monte Carlo move to transition to the $\xi_j$-sector is proposed, where $j$ is chosen uniformly from $[1,N_{\xi}]$, $N_{\xi}$ is the number of $\xi$ values in the ensemble. Alternatively, we can update the sector locally by randomly choosing $j$ to be either $i+1$ or $i-1$, and if the new $j$ is outside $[1,N_{\xi}]$ the current sector does not change. In the following, we use uniform sampling to update the sector. The move is accepted with probability
\begin{equation}
A_{\xi-Trans}=\min\{1,(\frac{\xi_j}{\xi_i})^{N_p}\}.
\end{equation}
It is easy to see that detailed balance is satisfied for the $\xi$-Translate move. However, care must be taken when transitioning from a sector with $\xi\neq 0$ to the $\xi=0$ sector. That is, if the current $N_p$ is nonzero, any such move must be rejected since in the $\xi=0$ sector $N_p$ is always 0. Otherwise, if $N_p=0$, the transition to the $\xi=0$ sector is always accepted.
\par
While the resulting $\xi$-ensemble worm algorithm works, there is a problem with the $\xi$-Translate move, namely, the fact that after thermal equilibrium is reached, the ratios between different $\xi$-sectors ($a_i$ in Eq. (\ref{Zen})) are not equal. That is, some $\xi$-sectors will be sampled more often compared to others, and that is detrimental to the overall sampling efficiency. To deal with this problem, we combine the $\xi$-Translate move with Wang-Landau (WL) algorithm\cite{wang2001effi} to ensure every $\xi$-sector is being evenly sampled. The procedure is outlined below.
\par
\textit{Initialization}. A vector of factors $S(\xi_i)$ for $\xi$-ensemble is initialized to 0, also a histogram of visited $\xi$-sectors, $H(\xi_i)$, is created and it is initially 0. At first we set $f=1$.
\par
\textit{$\xi$-Translate (WL)}. This Monte Carlo move to update $\xi$-sector remains the same, but with WL algorithm\cite{wang2001effi}, the acceptance probability is modified as
\begin{equation}
A_{\xi-Trans-WL}=\min\{1,e^{S(\xi_i)-S(\xi_j)}(\frac{\xi_j}{\xi_i})^{N_p}\}.
\end{equation}
Finally, to ensure even sampling across $\xi$-ensemble, we need to update variables in the WL algorithm after all Monte Carlo steps.
\par
\textit{Post update}. After every Monte Carlo update (not just the $\xi$-Translate move), we increment $S(\xi_i)$ and $H(\xi_i)$ in the current $\xi_i$-sector:
\begin{equation}
S(\xi_i)\leftarrow S(\xi_i)+f,\ H(\xi_i)\leftarrow H(\xi_i)+1.
\end{equation}
Then we check the flatness of the histogram $H(\xi)$, if it is flat enough, we decrease the value of $f$ by $f\leftarrow f/2$, and reset all $H(\xi)$ to be 0. With the WL version of the algorithm, we can expect that after equilibrium is reached, the samples will be evenly distributed across each $\xi$-sector, leading to roughly the same statistical fluctuation for all sectors in the $\xi$-ensemble. Generally speaking, the frequency of the $\xi$-Translate move should be set appropriately (usually lower than the other types of Monte Carlo moves), so that the autocorrelation time is minimized.
\par
The above paragraphs present the modifications for worm algorithm that is needed to take $\xi$-ensemble into account. Now, in order to analyze the statistical efficiency of the modified Monte Carlo algorithm, we use autocorrelation time as the main criterion. Specifically, we collect all the samples generated from Monte Carlo simulation that belong to the same $\xi$-sector, calculate some observable $O$ for those samples, and a set of values $\{O^{\xi_i}_j\}$ in the $\xi_i$-sector is obtained. Next, the unnormalized autocorrelation function $C_{\xi_i}(k)$ for the observable $O$ is defined as
\begin{equation}
C_{\xi_i}(k)=\frac{1}{M_{\xi_i}-k}\sum_{j=1}^{M_{\xi_i}-k}(O^{\xi_i}_j-\langle O^{\xi_i}\rangle)(O^{\xi_i}_{j+k}-\langle O^{\xi_i}\rangle).
\end{equation}
The autocorrelation function $C_{\xi_i}(k)$ decays with displacement $k$, and the autocorrelation time is about the half width of its distribution. Alternatively, it is more precise to use integrated autocorrelation time as the main measure, which is defined as
\begin{equation}
C^{\text{int}}_{\xi_i}=1+2\sum_{k=1}^{k_{max}}\frac{C_{\xi_i}(k)}{C_{\xi_i}(0)}.
\label{cint}
\end{equation}
In the following, we perform some simulations to verify that the autocorrelation time for $\xi$-ensemble is reduced compared to the case where we are doing simulation for each $\xi$-sector independently. It is common to consider autocorrelation for energy and centroid as the observables, where we define the centroid for the system as 
\begin{equation}
\mathbf{r}^c=\frac{1}{NP}\sum_{i=1}^N\sum_{j=1}^P\mathbf{R}_j^i.
\end{equation}

\section{Results}
In this section, simulation results for uniform electron gas\cite{dornheim2018uniform} and uniform warm dense beryllium\cite{dornheim2025unraveling,doppner2023observing} are shown. We follow the usual practice and use $\theta$ as temperature parameter and $r_s$ as density parameter, where $\theta$ is defined in terms of Fermi energy $E_F$ as
\begin{equation}
\theta=\frac{k_BT}{E_F},
\end{equation}
whereas $r_s$ is defined using electron density $n$
\begin{equation}
r_s=\left(\frac{3}{4\pi n}\right)^{1/3}.
\end{equation}
In the following we choose $\theta$ and $r_s$ so that the system is at warm dense matter condition, and isothermal $\xi$-extrapolation\cite{xiong2022thermodynamic,dornheim2023fermionic} can be used to compare against previous benchmarks.

\subsection{Uniform electron gas}
The model Hamiltonian defining uniform electron gas is given by\cite{dornheim2018uniform}
\begin{equation}
\hat H=-\frac{1}{2}\sum_{i=1}^N\nabla_i^2+\sum_{i<j}^NW_E(\mathbf{r}_i,\mathbf{r}_j)+\frac{N}{2}\xi_M,
\end{equation}
where $\xi_M$ is the Madelong constant, and $W_E$ is the Ewald pair potential defined as
\begin{equation}
W_E(\mathbf{r},\mathbf{s})=\frac{1}{V\pi}\sum_{\mathbf{G}\neq 0}(G^{-2}e^{-\frac{\pi^2G^2}{\kappa^2}+2\pi i\mathbf{G}\cdot(\mathbf{r}-\mathbf{s})})-\frac{\pi}{\kappa^2V}+\sum_{\mathbf{R}}\frac{\text{erfc}(\kappa|\mathbf{r}-\mathbf{s}+\mathbf{R}|)}{|\mathbf{r}-\mathbf{s}+\mathbf{R}|},
\end{equation}
where $V=L^3$ is the volume of the system, $\mathbf{G}=\mathbf{n}L^{-1}$, $\mathbf{R}=\mathbf{m}L$ for integer vectors $\mathbf{n}$ and $\mathbf{m}$. Here we consider two component electrons in the fully unpolarized case, so that $N_{\uparrow}=N_{\downarrow}=N/2$.
\par

\begin{figure}[htbp]
\begin{center}
\includegraphics[width=0.7\textwidth]{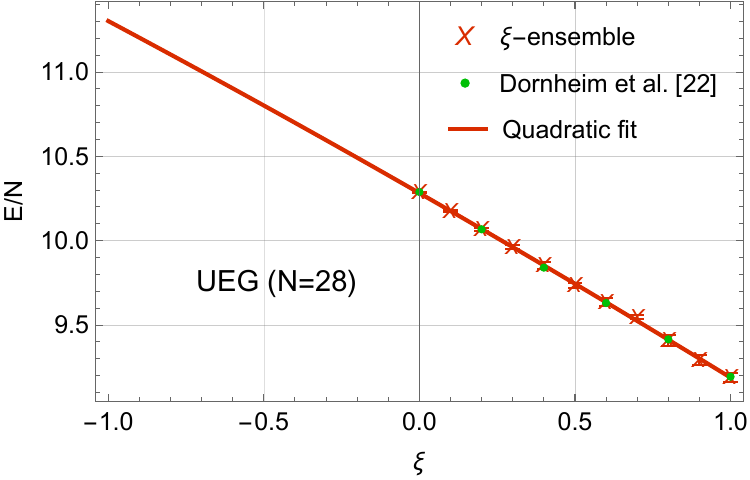}
\caption{\label{UEGN28E} The average energy of the uniform electron gas at $r_s=0.5$ and $\theta=1.0$, for 28 electrons. The red crosses represent the results obtained from the $\xi$-ensemble PIMC method, while the green dots denote the previous result reported in Ref.~\cite{dornheim2023fermionic}. The red solid line shows a quadratic fit to the $\xi$-ensemble PIMC data. 15 independent $\xi$-ensemble PIMC simulations each with $1.5\times 10^7$ Monte Carlo steps are used to calculate the average energy. }
\end{center}
\end{figure}

First, to validate the $\xi$-ensemble algorithm, the average energy of the system for 28 electrons at $r_s=0.5$ and $\theta=1.0$ is calculated at different $\xi$ values in the bosonic sector. For this system, the $\xi$-Translate move is applied once every 50 Monte Carlo steps. The results, along with isothermal $\xi$-extrapolation, are shown in Fig. \ref{UEGN28E}, and we found excellent agreement with previous benchmark\cite{dornheim2023fermionic} by Dornheim et al., showing the correctness of the algorithm and implementation.
\par

\begin{figure}[htbp]
\begin{center}
\includegraphics[width=1.0\textwidth]{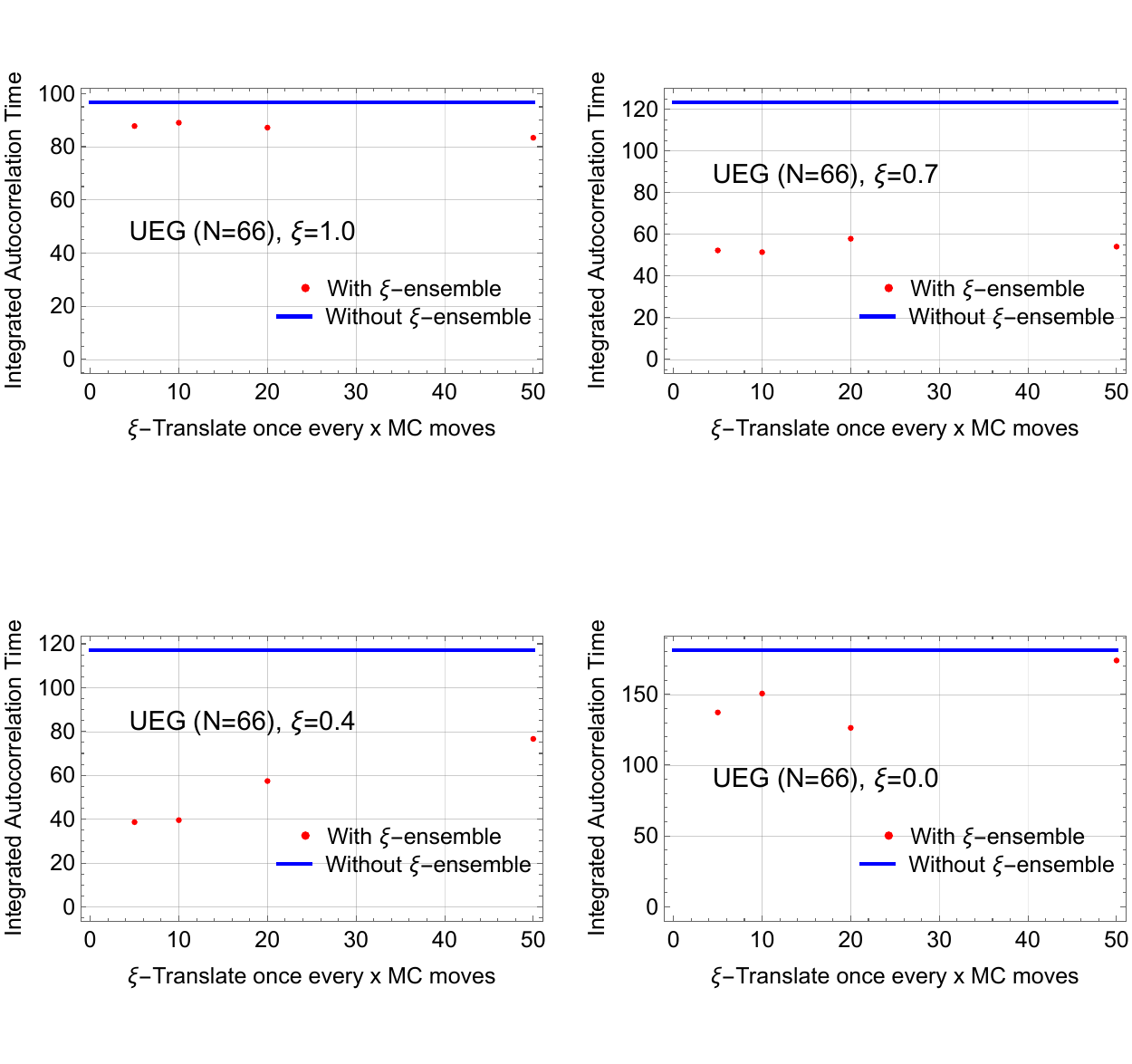}
\caption{\label{UEGN66Corr} The integrated autocorrelation time of energy, for uniform electron gas with 66 electrons. Selective results for 4 different $\xi$-sectors are shown: $\xi=1.0$ (top left panel), $\xi=0.7$ (top right panel), $\xi=0.4$ (bottom left panel) and $\xi=0.0$ (bottom right panel). Blue line is the autocorrelation time calculated from standard PIMC without $\xi$-ensemble. Red dot is the result for $\xi$-ensemble PIMC, where the $\xi$-Translate (WL) move is applied once every $x$ Monte Carlo steps. The reduction in autocorrelation time depends on specific $\xi$-sector and $\xi$-Translate move frequency. $10^6$ MC steps were used to generate the samples, and $k_{max}$ is chosen to be 150 when calculating integrated autocorrelation time. }
\end{center}
\end{figure}

Next, to show the improvement in sampling efficiency, we calculate the energy and centroid autocorrelation functions for the cases with and without using $\xi$-ensemble in PIMC. First, we investigate the integrated autocorrelation time (Eq. (\ref{cint})) for energy, for different frequencies of the $\xi$-Translate move. In this example, the $\xi$-ensemble consists of 11 different $\xi$-sectors, evenly spaced in $[0,1]$, and the integrated energy autocorrelation time 
for a few selected $\xi$-sectors are shown in Fig. \ref{UEGN66Corr}. We note that while the reduction of autocorrelation time depends on the $\xi$-sector and frequency of the $\xi$-Translate move, on average we can expect about $35\%$ reduction for this system.
\par

\begin{figure}[htbp]
\begin{center}
\includegraphics[width=1.0\textwidth]{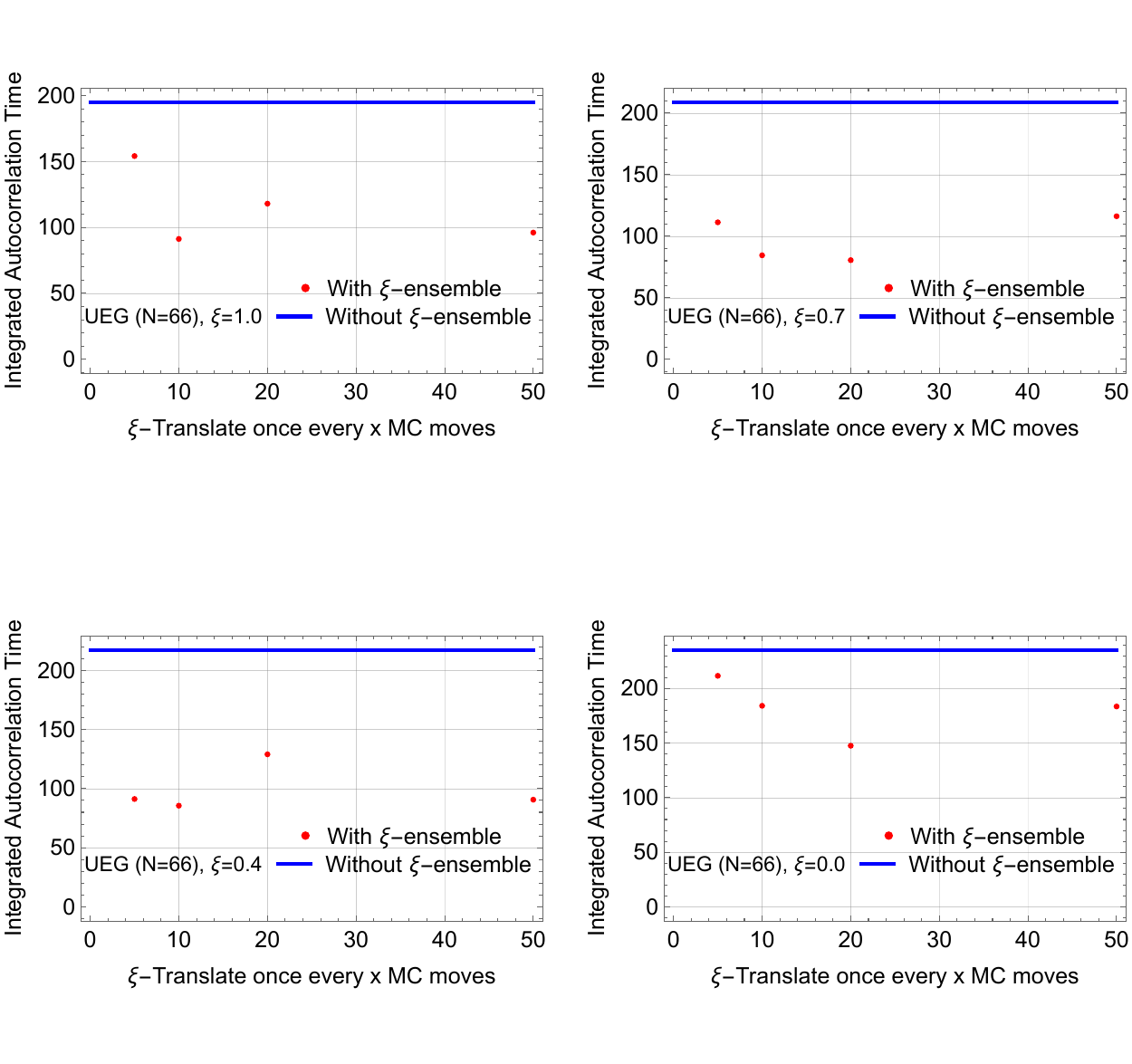}
\caption{\label{UEGN66CorrRc} The integrated autocorrelation time of $r^c_x$, for uniform electron gas with 66 electrons. Selective results for 4 different $\xi$-sectors are shown: $\xi=1.0$ (top left panel), $\xi=0.7$ (top right panel), $\xi=0.4$ (bottom left panel) and $\xi=0.0$ (bottom right panel). Blue line is the autocorrelation time calculated from standard PIMC without $\xi$-ensemble. Red dot is the result for $\xi$-ensemble PIMC, where the $\xi$-Translate (WL) move is applied once every $x$ Monte Carlo steps. $10^6$ MC steps were used to generate the samples, and $k_{max}$ is chosen to be 150 when calculating integrated autocorrelation time. On average there is about $42\%$ reduction in autocorrelation time compared to standard PIMC. }
\end{center}
\end{figure}

\begin{figure}[htbp]
\begin{center}
\includegraphics[width=0.7\textwidth]{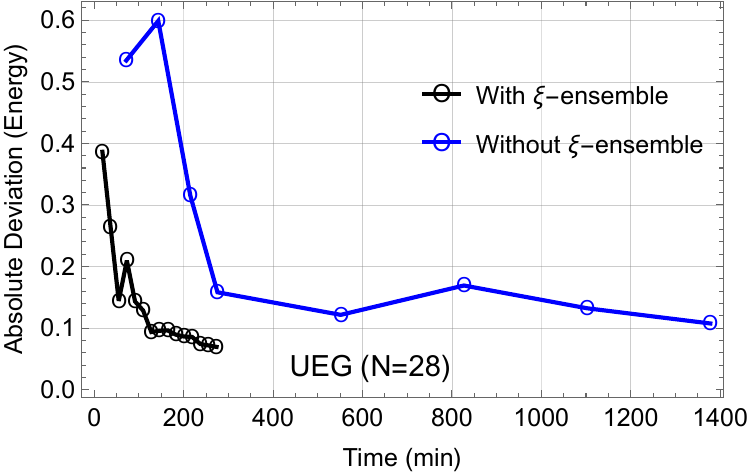}
\caption{\label{UEGN28Time} The total deviation of energy as a function of computational time, for uniform electron gas with 28 electrons at $r_s=0.5$ and $\theta=1.0$. We define the total deviation as the sum of the absolute deviations of energy for 11 $\xi$-sectors evenly spaced between $[0,1]$ and previous benchmark\cite{dornheim2023fermionic}. Black circles and blue circles are the simulation results obtained for $\xi$-ensemble PIMC and standard PIMC algorithms, respectively. The convergence of deviation to 0 is much faster for $\xi$-ensemble PIMC. }
\end{center}
\end{figure}

The integrated autocorrelation time for centroid is also calculated, and some selected results are shown in Fig. \ref{UEGN66CorrRc}. Similar to energy, the correlation time is not even across different $\xi$-sectors, and on average there is about $42\%$ reduction in autocorrelation time for centroid.
\par
Finally, to compare the performance of standard PIMC and $\xi$-ensemble PIMC more directly, we also plot the deviation as a function of computational time, for both algorithms, as shown in Fig \ref{UEGN28Time}. We note that, compared to standard PIMC, the convergence of deviation with respect to actual computational time is much faster when using $\xi$-ensemble PIMC. This also shows that in practice, if we use the $\xi$-ensemble PIMC algorithm we can save even more computational time than what the reduction in autocorrelation time may imply. This is because when we are sampling over $\xi$-ensemble, the different permutation groups can be more easily sampled compared to standard PIMC, and so in the same amount of computational time more statistically independent samples can be generated, and this improves the overall sampling efficiency.

\subsection{Uniform warm dense beryllium}
The model Hamiltonian defining uniform beryllium system is given by\cite{dornheim2024ab}
\begin{equation}
\begin{split}
\hat H=&-\frac{1}{2}\sum_{i=1}^{N_e}(\mathbf{\nabla}_i^e\cdot\mathbf{\nabla}_i^e)-\frac{1}{2m_b}\sum_{j=1}^{N_b}(\mathbf{\nabla}_j^b\cdot\mathbf{\nabla}_j^b)+\sum_{i<j}^{N_e}[W_E(\mathbf{r}_i,\mathbf{r}_j)-\xi_M]+16\sum_{i<j}^{N_b}[W_E(\mathbf{I}_i,\mathbf{I}_j)-\xi_M]\\
&-4\sum_{i=1}^{N_e}\sum_{j=1}^{N_b}[W_E(\mathbf{r}_i,\mathbf{I}_j)-\xi_M],
\end{split}
\end{equation}
where $\mathbf{\nabla}_i^e$ and $\mathbf{\nabla}_j^b$ act on the electron and beryllium ion coordinates, respectively. $m_b$ is the mass of ion, and $\mathbf{r}$ and $\mathbf{I}$ denote the electron and ion coordinates. $N_e$ and $N_b$ are the number of electrons and ions, and we choose $N_e=4N_b$ so that the system is charge neutral. In order to deal with the attractive Coulomb interaction between electrons and ions, pair approximation\cite{militzer2016comp,maximilian2023ab} is used.
\par

\begin{figure}[htbp]
\begin{center}
\includegraphics[width=1.0\textwidth]{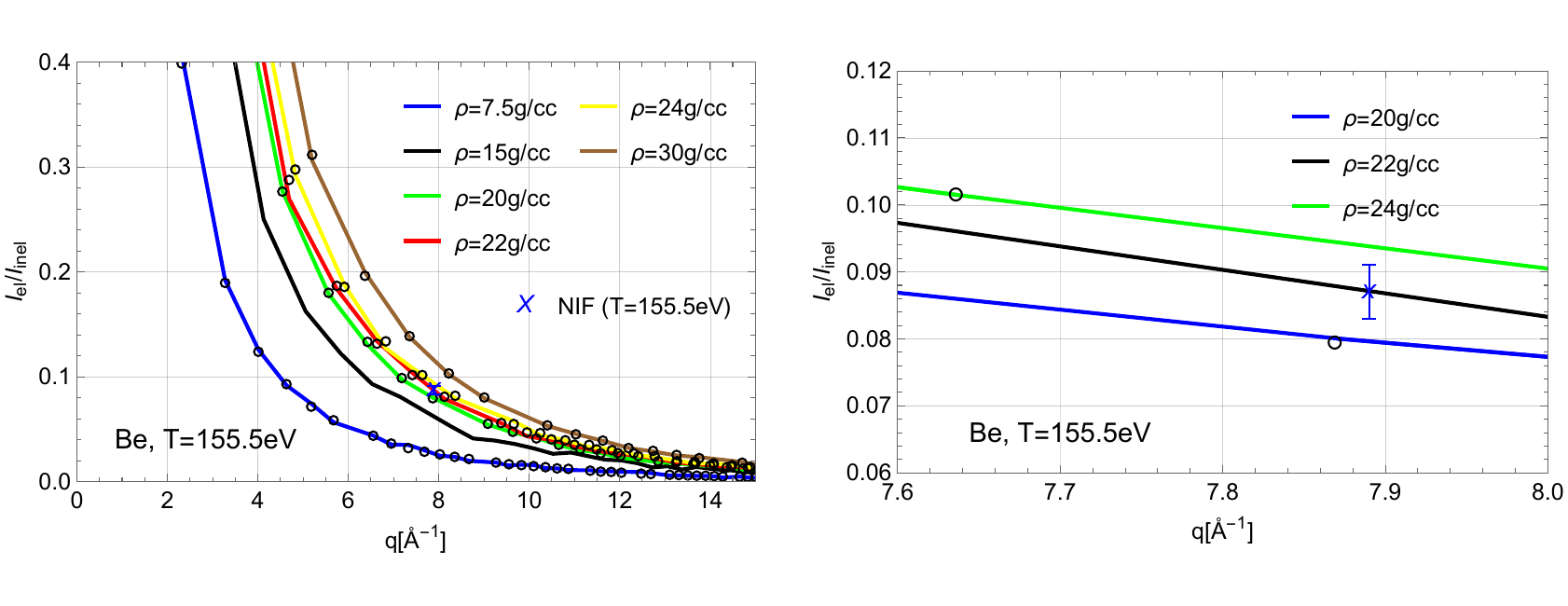}
\caption{\label{UBG155ev} $\frac{I_{{el}}(q)}{I_{{inel}}(q)}$ for 10 beryllium atoms in a periodic box at $T=155.5\text{eV}$ for a few different densities. The solid lines are results calculated using $\xi$-ensemble and isothermal $\xi$-extrapolation. Black circles indicate previous PIMC simulations \cite{dornheim2025unraveling}, and blue crosses represent experimental results \cite{dornheim2025unraveling}. The right figure is an enlarged view of the left figure near National Ignition Facility (NIF) data point.}
\end{center}
\end{figure}

\begin{figure}[htbp]
\begin{center}
\includegraphics[width=1.0\textwidth]{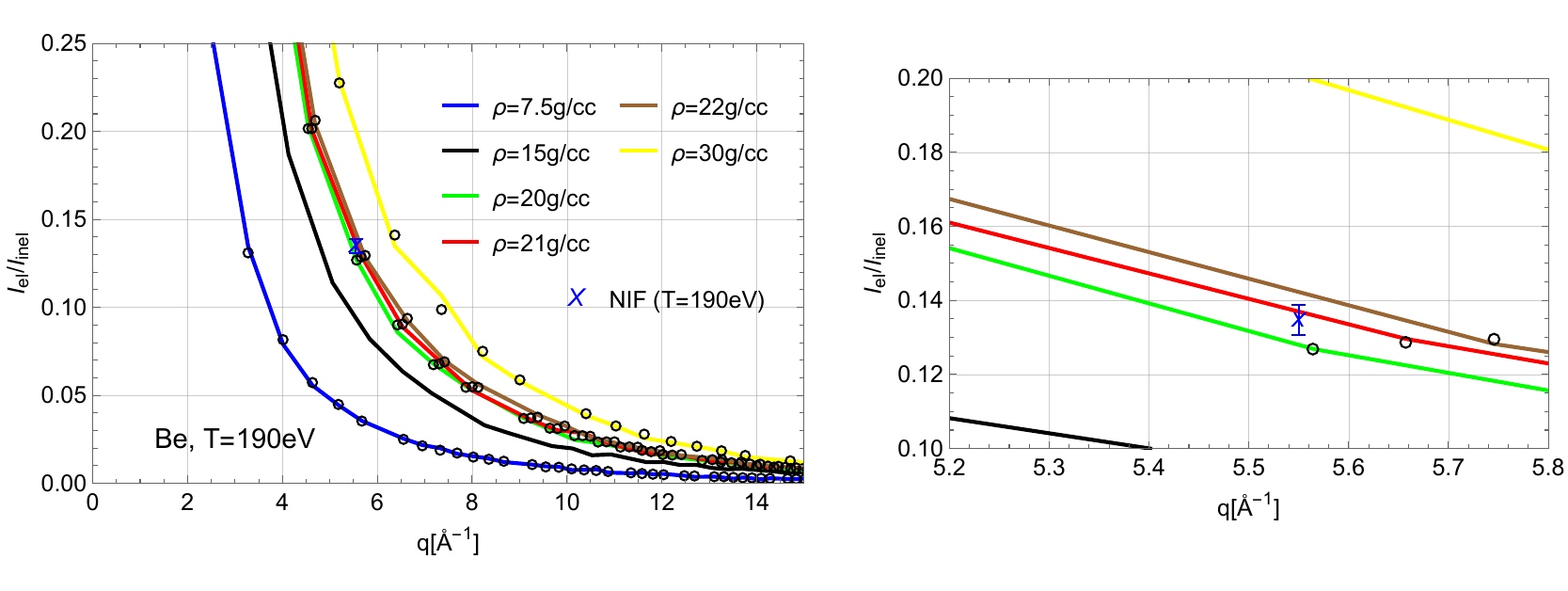}
\caption{\label{UBG190ev} $\frac{I_{{el}}(q)}{I_{{inel}}(q)}$ for 10 beryllium atoms in a periodic box at $T=190\text{eV}$ for a few different densities. The solid lines are results calculated using $\xi$-ensemble and isothermal $\xi$-extrapolation. Black circles indicate previous PIMC simulations \cite{dornheim2025unraveling}, and blue crosses represent experimental results \cite{dornheim2025unraveling}. The right figure is an enlarged view of the left figure near National Ignition Facility (NIF) data point. }
\end{center}
\end{figure}

Again, to test the correctness of implementation, simulation for 10 beryllium atoms in a periodic box at warm dense matter condition is performed. We used $\xi$-ensemble PIMC to calculate the observable $\frac{I_{{el}}(q)}{I_{{inel}}(q)}$, which is the ratio of elastic and inelastic contribution to the static structure factor, and is defined as\cite{dornheim2025unraveling}
\begin{equation}
\frac{I_{{el}}(q)}{I_{{inel}}(q)}=\left(\frac{S_{{ee}}(q)S_{{II}}(q)}{S_{eI}^2(q)}-1\right)^{-1},
\end{equation}
where $S_{{ee}}$, $S_{{II}}$ and $S_{eI}$ are the electron-electron, ion-ion and electron-ion static structure factors, respectively. In the bosonic sector, eleven $\xi$ values evenly spaced in the interval $[0,1]$ are chosen for $\xi$-ensemble, and isothermal $\xi$-extrapolation with quadratic fitting is used to calculate the observable for fermions. We apply the $\xi$-Translate move once every 50 Monte Carlo steps. The result of $\frac{I_{{el}}(q)}{I_{{inel}}(q)}$ is shown in Fig. \ref{UBG155ev} (for $T=155.5\text{eV}$) and Fig. \ref{UBG190ev} (for $T=190\text{eV}$), for a few different densities, and we found excellent agreement with previous theoretical and experimental results\cite{dornheim2025unraveling}.
\par

\begin{figure}[htbp]
\begin{center}
\includegraphics[width=1.0\textwidth]{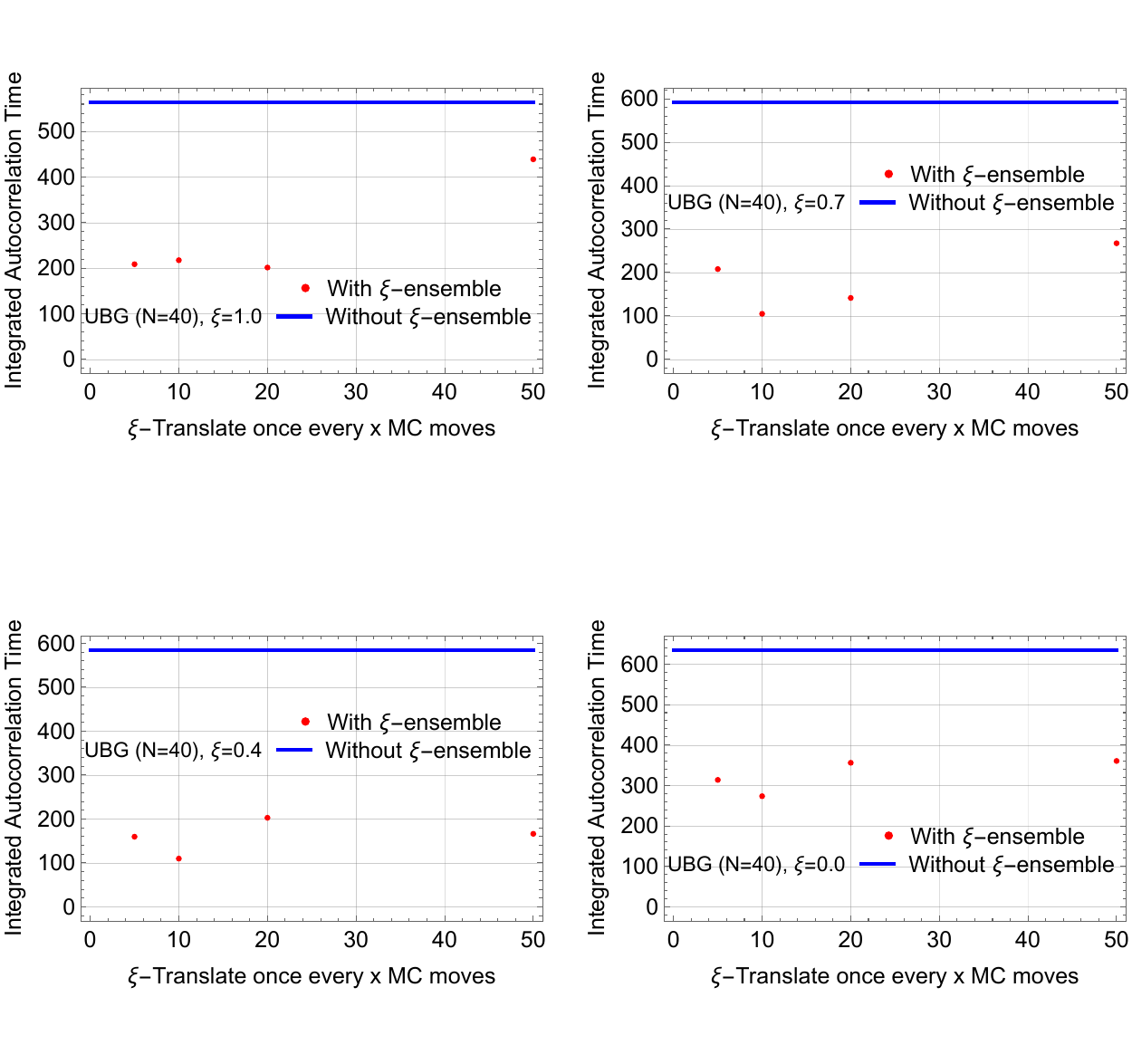}
\caption{\label{UBGN40Corr} The integrated autocorrelation time of energy, for uniform beryllium gas with 40 electrons. Selective results for 4 different $\xi$-sectors are shown: $\xi=1.0$ (top left panel), $\xi=0.7$ (top right panel), $\xi=0.4$ (bottom left panel) and $\xi=0.0$ (bottom right panel). Blue line is the autocorrelation time calculated from standard PIMC without $\xi$-ensemble. Red dot is the result for $\xi$-ensemble PIMC, where the $\xi$-Translate (WL) move is applied once every $x$ Monte Carlo steps. $10^6$ MC steps were used to generate the samples, and $k_{max}$ is chosen to be 500 when calculating integrated autocorrelation time. On average there is about $60\%$ reduction in autocorrelation time. }
\end{center}
\end{figure}

\begin{figure}[htbp]
\begin{center}
\includegraphics[width=1.0\textwidth]{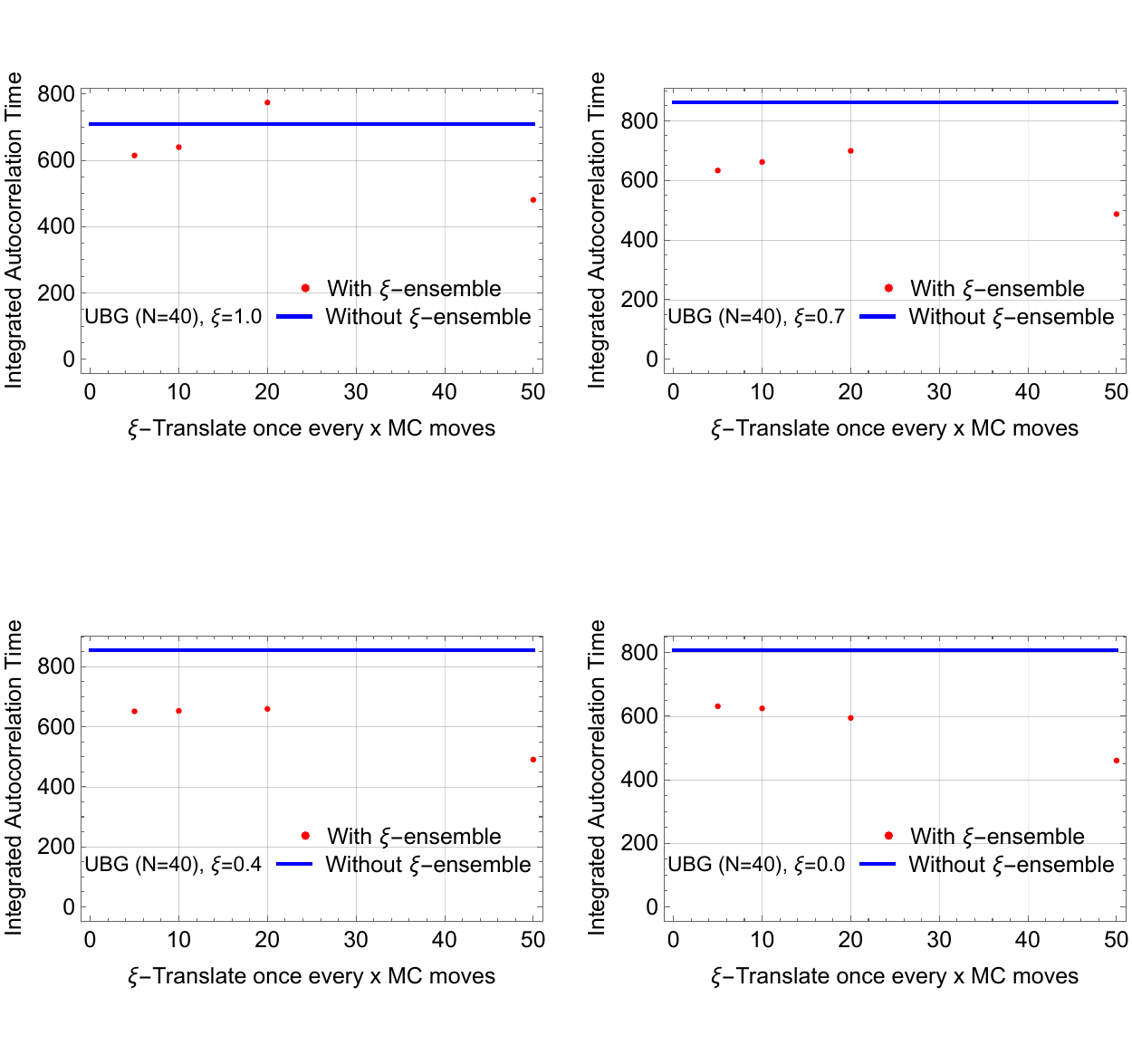}
\caption{\label{UBGN40CorrRc} The integrated autocorrelation time of $r^c_x$, for uniform beryllium gas with 40 electrons. Selective results for 4 different $\xi$-sectors are shown: $\xi=1.0$ (top left panel), $\xi=0.7$ (top right panel), $\xi=0.4$ (bottom left panel) and $\xi=0.0$ (bottom right panel). Blue line is the autocorrelation time calculated from standard PIMC without $\xi$-ensemble. Red dot is the result for $\xi$-ensemble PIMC, where the $\xi$-Translate (WL) move is applied once every $x$ Monte Carlo steps. $10^6$ MC steps were used to generate the samples, and $k_{max}$ is chosen to be 500 when calculating integrated autocorrelation time. On average there is about $24\%$ reduction in autocorrelation time. }
\end{center}
\end{figure}

To investigate the improvement in sampling efficiency for this system, we again calculate the energy and centroid autocorrelation functions with and without using $\xi$-ensemble. Some selected results are shown in Fig. \ref{UBGN40Corr} (for energy) and Fig. \ref{UBGN40CorrRc} (for centroid). Because of the longer autocorrelation time, a larger $k_{max}$ is used to calculate Eq. (\ref{cint}). For this system, we observed about $60\%$ and $24\%$ reduction in autocorrelation time on average, for energy and centroid respectively.
\par

\begin{figure}[htbp]
\begin{center}
\includegraphics[width=0.7\textwidth]{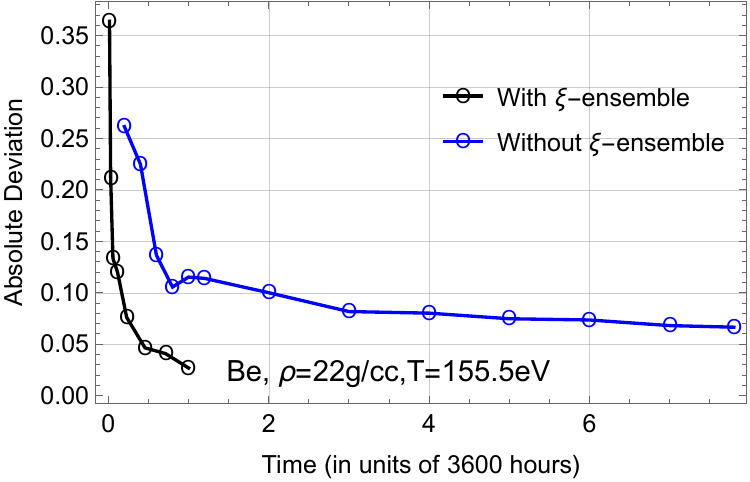}
\caption{\label{UBGN40Time} The deviation of $\frac{I_{{el}}(q)}{I_{{inel}}(q)}$ as a function of computational time, for uniform warm dense beryllium with 10 atoms at $\rho=22\text{g/cc}$ and $T=155.5\text{eV}$. The absolute deviation is calculated as the sum of the difference with previous benchmark\cite{dornheim2025unraveling} for all $q$ points. Black circles and blue circles are the simulation results obtained for $\xi$-ensemble PIMC and standard PIMC algorithms, respectively. The convergence of deviation to 0 is much faster for $\xi$-ensemble PIMC. }
\end{center}
\end{figure}

For this example, we also compare the performance of $\xi$-ensemble PIMC and standard PIMC directly, by plotting the deviation of $\frac{I_{{el}}(q)}{I_{{inel}}(q)}$ as a function of computational time, as shown in Fig. \ref{UBGN40Time}. Again, we can clearly see that compared to standard PIMC, $\xi$-ensemble PIMC needs much less computational time to reach a certain simulation accuracy.


\section{Summary}
As a summary, in this work a PIMC algorithm for fictitious identical particles is proposed, by combining worm algorithm with $\xi$-ensemble. This way, in a single PIMC simulation, properties for different $\xi$-sectors can be obtained at once. Tests on uniform electron gas and uniform warm dense beryllium are also performed to validate the algorithm and show its improvement in sampling efficiency, through analysis of autocorrelation time for different observables and convergence of deviation with computational time. We believe that $\xi$-ensemble PIMC will be useful for future applications of fictitious identical particles, to accelerate the simulation and improve its sampling efficiency. Moreover, it is also possible to combine the $\xi$-ensemble PIMC in this work, with other modified PIMC algorithms\cite{karmakar2026combining,xun2026exchange}, in order to further improve the efficiency of PIMC simulations.

\section*{Acknowledgments}
Y. Xiong gratefully acknowledges the support of the Hubei Provincial Young Top-Talent Program for this work. The author thanks H. Xiong for useful discussions.

\bibliography{biblio}

\end{document}